\documentclass[conference]{IEEEtran}
\IEEEoverridecommandlockouts
\usepackage{cite}
\usepackage{amsmath,amssymb,amsfonts}
\usepackage{algorithmic}
\usepackage{multirow}
\usepackage{graphicx}
\usepackage{textcomp}
\usepackage{xcolor}

\def\BibTeX{{\rm B\kern-.05em{\sc i\kern-.025em b}\kern-.08em
    T\kern-.1667em\lower.7ex\hbox{E}\kern-.125emX}}
\begin{document}

\hyphenation{Toyodabayashi}

\title{Analysis of Bias in Gathering Information Between User Attributes in News Application}

\author{\IEEEauthorblockN{Yoshifumi Seki}
\IEEEauthorblockA{\textit{Gunosy Inc.} \\
Tokyo, Japan \\
yoshifumi.seki@gunosy.com}
\and
\IEEEauthorblockN{Mitsuo Yoshida}
\IEEEauthorblockA{\textit{Toyohashi University of Technology}\\
Aichi, Japan \\
yoshida@cs.tut.ac.jp}
}

\maketitle

\begin{abstract}
In the process of information gathering on the web, confirmation bias is known to exist, exemplified in phenomena such as echo chambers and filter bubbles.
Our purpose is to reveal how people consume news and discuss these phenomena.
In web services, we are able to use action logs of a service to investigate these phenomena.
However, many existing studies about these phenomena are conducted via questionnaires, and there are few studies using action logs.
In this paper, we attempt to discover biases of information gathering due to differences in user demographic attributes, such as age and gender, from the behavior log of the news distribution service.
First, we summarized the actions in the service for each user attribute and showed the difference of user behavior depending on the attributes.
Next, the degree of correlation between the attributes was measured using the correlation coefficient, and a strong correlation was found to exist in the browsing tendency of the news articles between the attributes.
Then, the bias of keywords between attributes was discovered, keywords with bias in behavior among the attributes were found using parameters of regression analysis.
Since these discovered keywords are almost explainable by big news, our proposed method is effective in detecting biased keywords.
\end{abstract}

\begin{IEEEkeywords}
user behavior analysis, computation social science, news recommender system
\end{IEEEkeywords}

\section{Introduction}

In recent years, most people have been using the web to gather information.
One of differences between the web and the mass media is that, in the web, not everyone receives the same information; rather, individuals may access different information.
One of the most significant information-gathering behaviors on the web is {\it search}.
In search engines like Google\footnote{https://google.com/}, even if users search for the same keyword, what is clicked on the search results varies from user to user, and the search results differ according to the user's behavior\cite{Pariser2011}.
The Social Networking Services (SNSs) have spread rapidly in recent years.
In SNSs, a user receives information from the users and media one follows, so users receive different information.
In this way, as the web has become the center of information gathering, people can obtain various types of information according to their interests.

Under the present circumstances, problems referred to as {\it echo chambers} and {\it filter bubbles} have been pointed out\cite{Pariser2011, Jamieson2008}.
An echo chamber is a phenomenon in which an opinion is strengthened and amplified by repeating communication in a community where many people of the same opinion gather.
In an echo chamber situation, people tend to ignore wider opinion and exhibit a biased perspective.
This phenomenon became especially noticeable with the popularization of SNSs; for example, it had a significant influence on Donald Trump's victory in the 2016 US presidential election\cite{Hooton2016}.
A filter bubble, proposed by Eli Pariser in 2011\cite{Pariser2011}, is a problem that prevents the user from obtaining a wide range of information because search results, SNS posts, and recommended friends differ from one person to another due to the sophistication of the recommendation system.
Echo chambers and filter bubbles are similar in that they show that the information people can obtain is biased by the web.

These two problems are fundamentally based on a phenomenon called {\it selective exposure}, in which people find only the information they like\cite{Frey1986}.
Social surveys have analyzed these phenomena and how they are occurring\cite{Beam2014,Garrett2009}, and there are some studies that have looked at SNS data\cite{Barber2015}.
In terms of similar studies, there are studies that have extracted data from the SNS community using Twitter\footnote{https://twitter.com/} data.
However, 56\% of contents shared by SNSs are not being read\cite{Gabielkov2016}, and the motivations for sharing to SNSs are mainly {\it status seeking} and {\it socializing} \cite{Lee2012}.
For this reason, the SNS data don't reflect the real world enough to discuss these phenomena.
So we directly analyze user behavior to clarify the bias of the user.

Our purpose is to clarify the bias of user information gathering.
In this paper, we analyze how behavior in the news distribution service differs between user attributes, such as gender and age, and identify topics with bias between attributes.
Ideally, we would like to analyze users based on their interests, but for that purpose, discussion of the method of extracting and classifying users' interests is required.
The classification by demographic attributes reflects a user's interests to a certain extent, because such attributes are often used to solve the cold-start problem of the recommendation system\cite{Safoury2013}.
To focus on showing a method of clarifying the difference, we adopted the classification of demographic attributes as an interest classification method.

First, we confirm the correlation between the numbers of clicks and likes for each news article for each attribute and show how the user behavior between the attributes is different.
Next, we consider the keywords appearing in the title of the news article, thereby confirming the difference in the interest tendency of the news articles because there is a difference in the manner of correlation between the click and like attributes of each keyword.
Regression analysis is carried out for each keyword, and we attempt to clarify the bias of keywords from the parameters of the regression analysis.

\section{Dataset}

In this study, we employed user activity logs in a mobile application from August 1 to 31, 2018.
The data were gathered using Gunosy\footnote{https://gunosy.co.jp/en}, an information curation service for mobile applications.
Gunosy is a well-known mobile application in Japan, and it has been downloaded over 24 million times.

\begin{figure}[tp]
  \centering
  \includegraphics[width=0.99\linewidth]{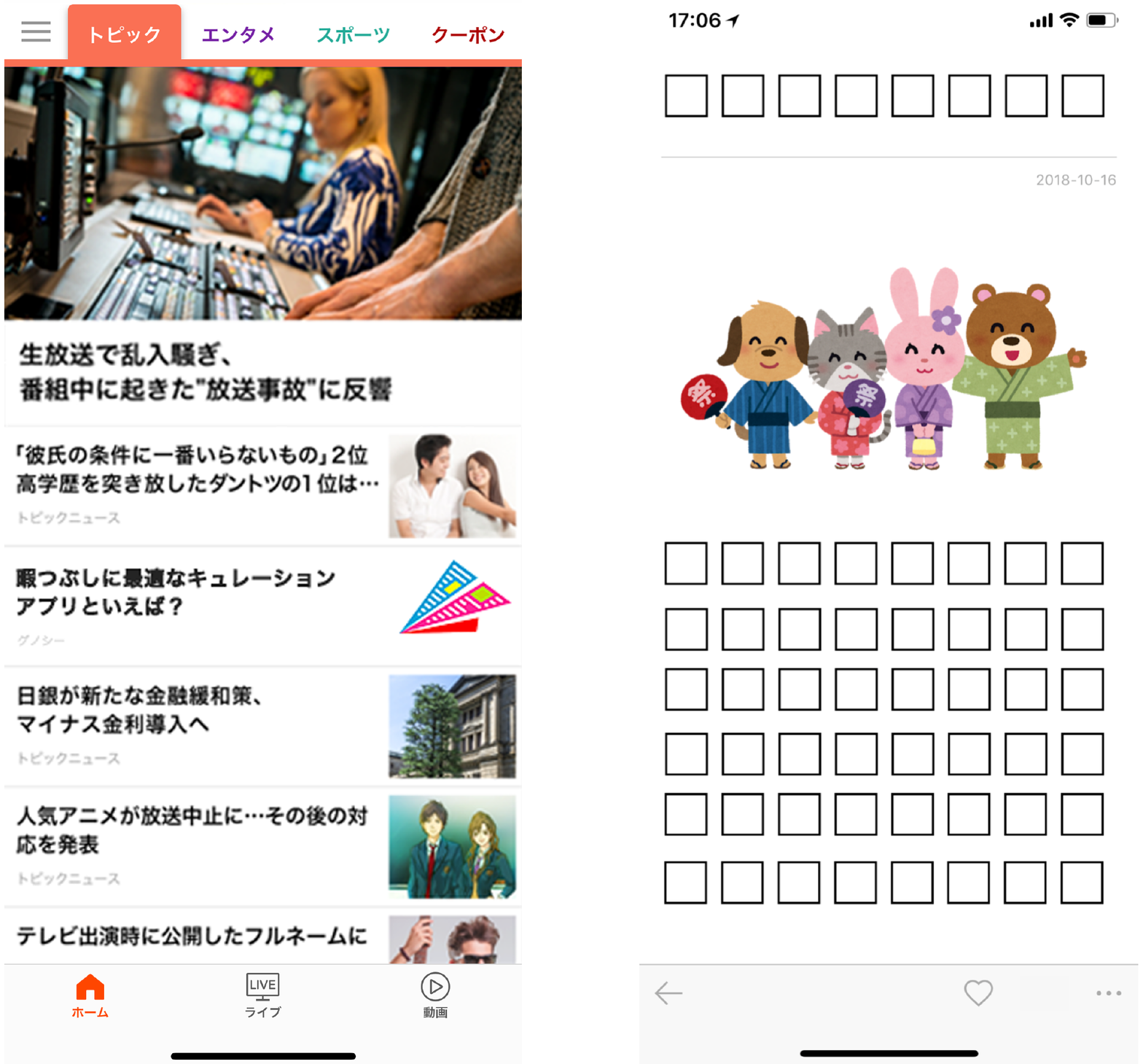}
    \caption{An example of news displayed on Gunosy: Left figure is a news list. A news list is displayed with photos, and detailed news content can be viewed by clicking the news title.
    Right figure is news article detail view.
In the bottom toolbar of this view, there is an icon of a hart as a {\it like} button.}
    \label{fig:gunosy}
\end{figure}

Fig.~\ref{fig:gunosy} shows how news is displayed on the application.
Various news articles are distributed in Gunosy, but in this research we use news articles in the categories of {\it politics} and {\it society}.
In Gunosy, news articles are classified into categories by several heuristic rules and supervised machine learning.

We use two types of user behavior for news articles.
First, we use the log of which news article the user clicked.
Users using this application can read news articles by clicking the cell of each news item, comprising a title and a thumbnail image.
Second, we use the log of which news article the user {\it liked}.
One can {\it like} news articles by clicking the heart icon on the toolbar at the bottom of the application or the heart icon at the end of the news article at the transition destination where one clicked the news article.
In some services, for example Twitter, {\it liked} tweets are stocked; however, in Gunosy, currently, the {\it like} does not provide a concrete benefit to a user.
But the {\it like} is used by some users.
We analyze how these two user behaviors differ depending on user attributes.

We use gender and age as user attributes.
The user registers these attributes at the start of using the service.
In Gunosy, gender is divided into three categories: male, female, and others.\footnote{For analysis, we use only male or female users.}
Age is divided into six categories; under 20, 20--24, 25--29, 30s, 40s, and over 50.
For analysis, these categories were converted into three broader categories: less than 30 years ({\it young}), 30s ({\it middle}), and 40s or more ({\it older}).
If the user doesn't register their attribute to our service, we predict their attributes using supervised machine learning and their behavior in applications.
Although some users aren't defined their attributes, we only use log data of those users whose attributes are defined.

We analyze article clicks/likes between user attributes.
Since {\it like} can only be done after clicking, article likes are much smaller than the article clicks.
Thus, the datasets consist of articles clicked more than 100 times.

\section{Comparing Actions between User Attributes}

\begin{table}[tp]
\centering
\caption{Each action ratio between demographic attributes.}
\begin{tabular}{|c|c|c|c|c|c|}
\hline
\multicolumn{3}{|l|}{}                                          & all    & \multicolumn{1}{l|}{Politics} & \multicolumn{1}{l|}{Society} \\ \hline
\multicolumn{3}{|c|}{number of news articles}                   &        & 1,333                         & 8,801                       \\ \hline
\multirow{5}{*}{Click ratio} & \multirow{2}{*}{Gender} & Male   & 58.9\% & 76.2\%                        & 54.0\%                      \\
                             &                         & Female & 41.1\% & 23.8\%                        & 46.0\%                      \\ \cline{2-6} 
                             & \multirow{3}{*}{Age}    & Young  & 34.7\% & 16.4\%                        & 23.1\%                      \\
                             &                         & Middle & 30.2\% & 22.1\%                        & 30.4\%                      \\
                             &                         & Older  & 35.1\% & 61.5\%                        & 46.5\%                      \\ \hline
\multirow{5}{*}{Like ratio}  & \multirow{2}{*}{Gender} & Male   & 47.7\% & 78.2\%                        & 47.4\%                      \\
                             &                         & Female & 52.3\% & 21.8\%                        & 52.6\%                      \\ \cline{2-6} 
                             & \multirow{3}{*}{Age}    & Young  & 25.8\% & 8.8\%                         & 16.0\%                      \\
                             &                         & Middle & 25.4\% & 11.0\%                        & 22.1\%                      \\
                             &                         & Older  & 48.7\% & 80.2\%                        & 61.9\%                      \\ \hline
\end{tabular}
\label{tbl:action_ratio}
\end{table}

Table~\ref{tbl:action_ratio}. shows the number of articles in the datasets and the composition ratio of each action for each attribute for each category in the whole application.
The composition ratio of each behavior differs greatly depending on the categories.
For example, males clicked more than females by a greater margin in politics than in society.
In addition, comparing likes with clicks, older people used {\it like} more than young people did.
Since the composition ratio of each behavior is different depending on the categories, there is certain validity in handling the attribute as a user's interest difference.

\begin{figure}[tp]
  \centering
  \includegraphics[width=0.99\linewidth]{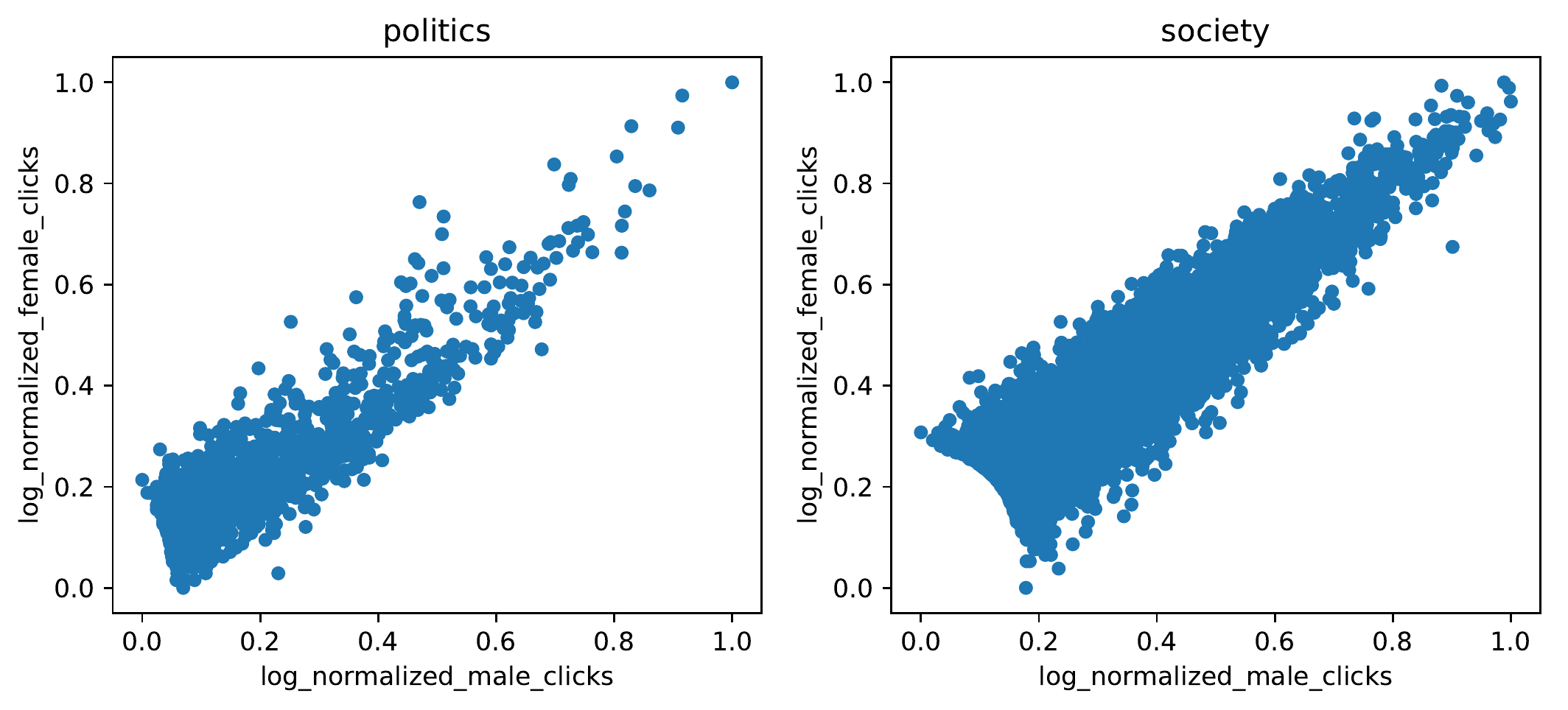}
    \caption{Scatter plots of the numbers of clicks on news articles by genders. }
    \label{fig:scatter_gender_clicks}
\end{figure}

\begin{figure}[tp]
  \centering
  \includegraphics[width=0.99\linewidth]{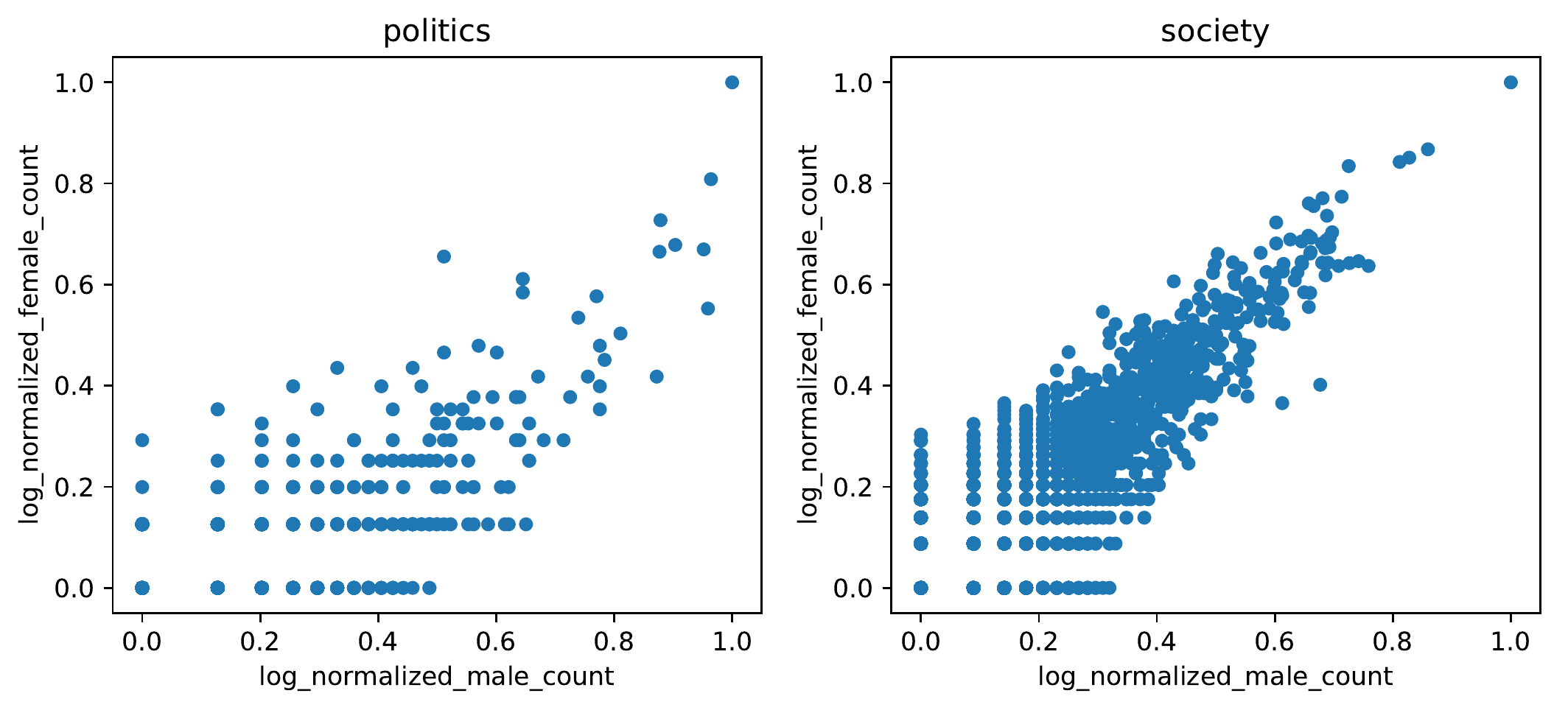}
    \caption{Scatter plots of the numbers of likes on news articles by genders. }
    \label{fig:scatter_gender_likes}
\end{figure}

Next, we show the relationships between attributes.
Because the composition ratio is different, the number of clicks and likes of articles in attributes and categories must be normalized.
The normalization was defined as $normalize(x) = \frac{log(x) - min(log(x))}{max(log(x)) - min(log(x))}$ where the number of clicks and the number of likes is $x$.

Fig.~\ref{fig:scatter_gender_clicks} shows the scatter plots of the normalized numbers of clicks in news articles by males and females, while Fig.~\ref{fig:scatter_gender_likes} shows the scatter plots of the normalized numbers of likes in news articles by males and females.
Pearson's correlation coefficient for the normalized numbers of clicks on news articles between male and female had very strong positive correlations of 0.902 for politics and 0.883 for society.
Pearson's correlation coefficient for the normalized numbers of likes between males and females is politics of 0.502 and society of 0.509, and this is weakly than click's one.
There was not a big difference in the relative clicks of news articles between males and females, but the difference in the relative likes of news articles between males and females was larger than the difference in clicks.

\begin{figure}[tp]
  \centering
  \includegraphics[width=0.99\linewidth]{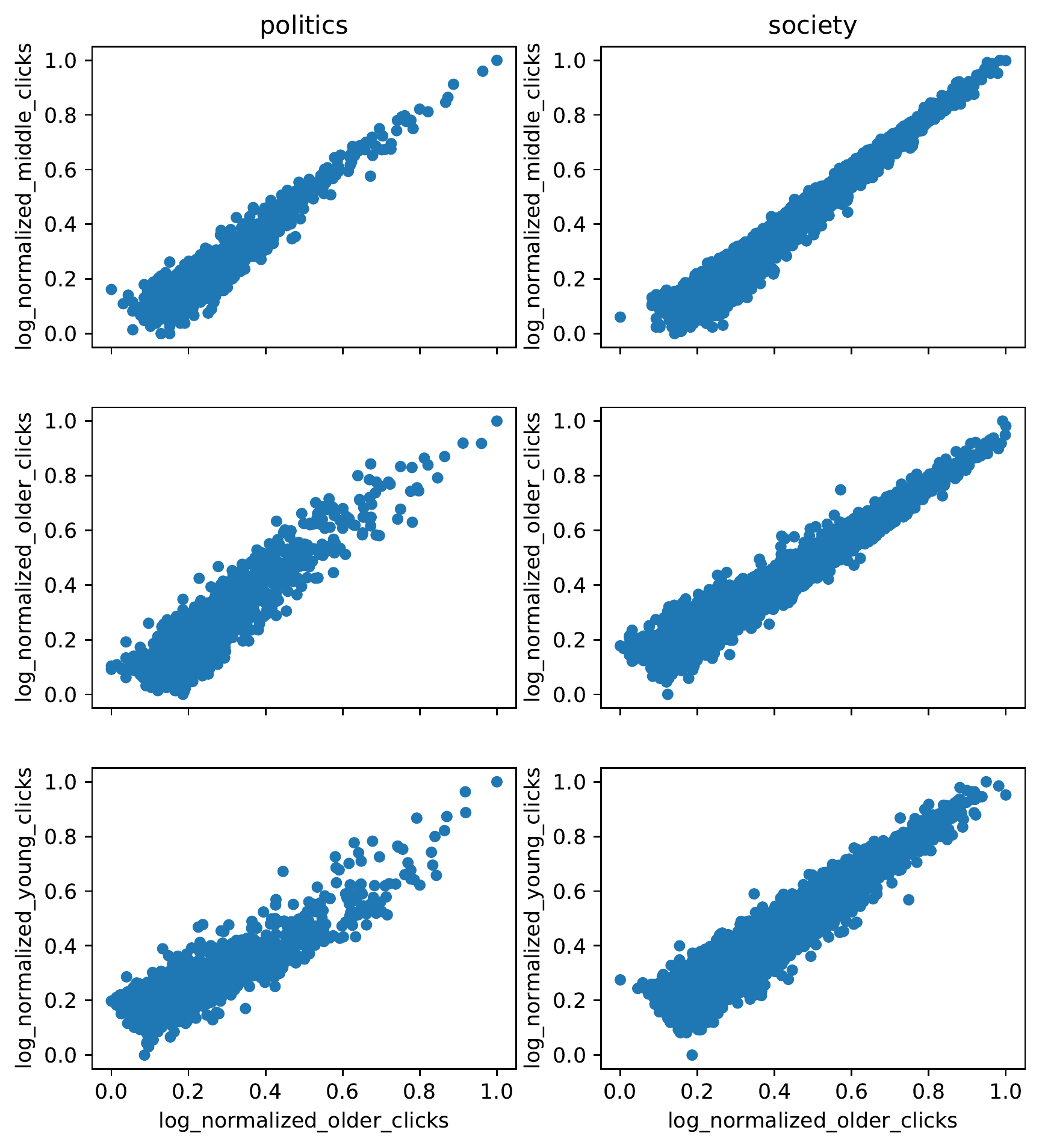}
    \caption{Scatter plots of the numbers of clicks on news articles by age.}
    \label{fig:scatter_age_clicks}
\end{figure}

\begin{figure}[tp]
  \centering
  \includegraphics[width=0.99\linewidth]{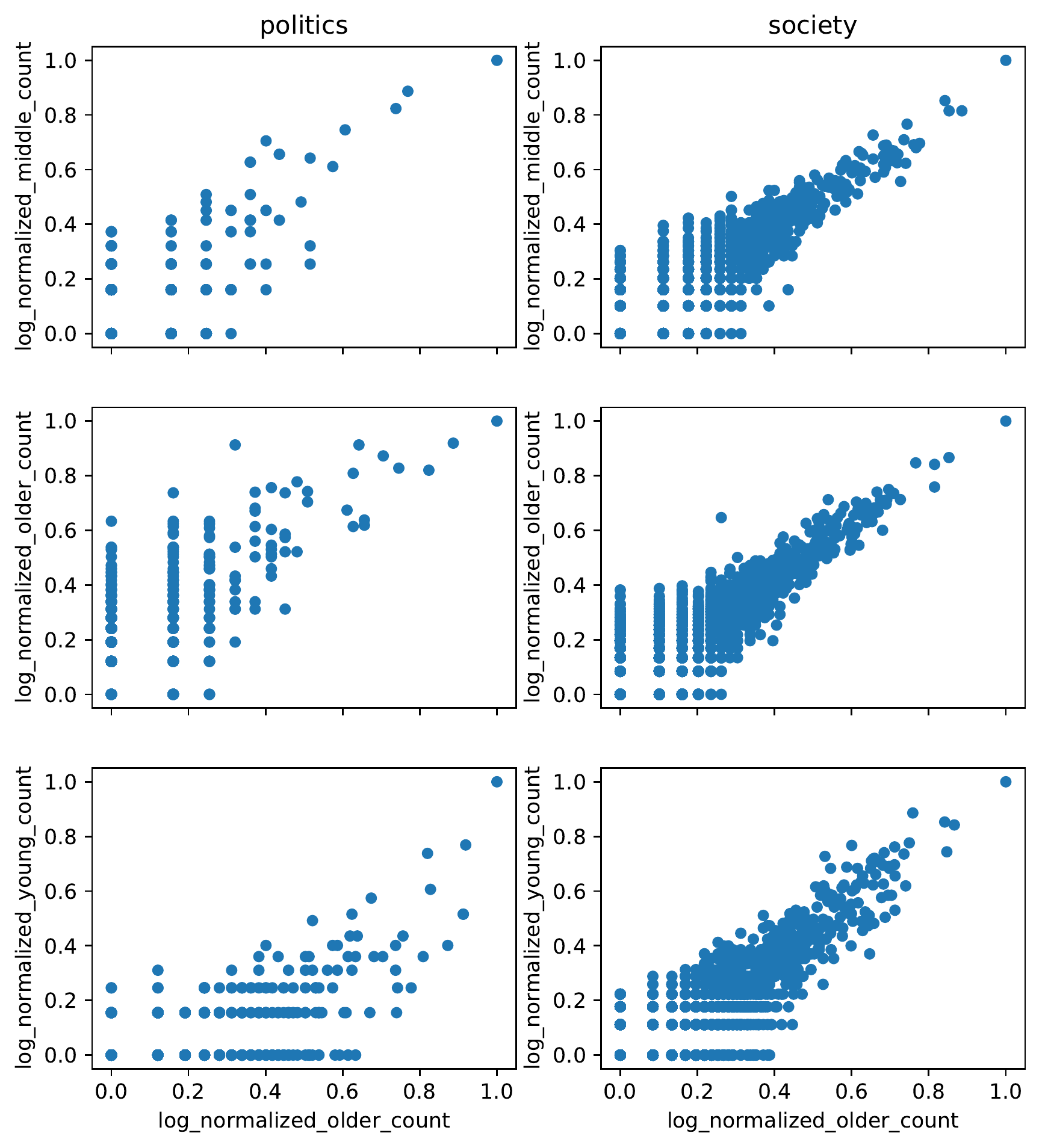}
    \caption{Scatter plots of the numbers of likes on news articles by age.}
    \label{fig:scatter_age_likes}
\end{figure}

Fig.~\ref{fig:scatter_age_clicks} shows the scatter plots of the normalized numbers of clicks on news articles by age, while Fig.~\ref{fig:scatter_age_likes} shows the scatter plots of the normalized numbers of likes on news articles by age.
Pearson's correlation coefficient of the young and middle-aged categories clicks for politics was 0.993, that of middle and older was 0.923, and that of older and young was 0.901.
For society, Pearson's correlation coefficient for the young and middle categories was 0.985, that for middle and older was 0.969, and that for older and young was 0.936.
The normalized number of clicks correlated very strongly, showing that the correlation was stronger for age than it was for gender.
Since the normalized number of clicks correlated very strongly, and since the coefficient for age was stronger than that for gender, the difference in interest in news articles is larger between genders than it is between ages.
For politics, Pearson's correlation coefficients for the young and middle categories in the normalized numbers of likes was 0.909, that for middle and older users was 0.845, and that for older and young people was 0.786.
In terms of society, Pearson's correlation coefficients for young and middle users in normalized number of likes was 0.955, that for middle and older people was 0.976, and that for older and young people was 0.902.
The normalized number of likes was also strongly correlated overall, so it was very strong compared with gender.
Politics showed a weaker correlation than society did, and a relatively weak correlation was exhibited between young and older people compared to young and middle people.

As for the overall trend, the correlation of normalized clicks was higher than that of normalized likes, but the latter was also strong.
In addition, the correlation for society tended to be stronger than the correlation for politics.
In the correlation of age, that between young people and middle-aged people was the strongest, while that between young people and older people was the weakest.
Since all the relationships had strong correlations, there is validity in comparing the relations between attributes using clicks and likes.

\begin{table}[tp]
\centering
\caption{Demographic Rank Correlation Coefficient.}
\scalebox{0.90}{
\begin{tabular}{|c|c|c|c|c|c|c|c|}
\hline
\multicolumn{2}{|c|}{\multirow{2}{*}{}} & \multicolumn{2}{c|}{Click} & \multicolumn{2}{c|}{Like} \\ \cline{3-6} 
\multicolumn{2}{|c|}{}                  & Politics     & Society     & Politics     & Society    \\ \hline
Gender                 & Male--Female    & 0.821        & 0.861       & 0.501        & 0.561            \\ \hline
\multirow{3}{*}{Age}   & Young--Middle   & 0.899        & 0.967       & 0.368        & 0.539             \\ \cline{2-6} 
                       & Middle--Older   & 0.849        & 0.958       & 0.433        & 0.556           \\ \cline{2-6} 
                       & Older--Young    & 0.818        & 0.932       & 0.370        & 0.532           \\ \hline
\end{tabular}
}
\label{tbl:rank_cor}
\end{table}

Table~\ref{tbl:rank_cor}. shows Spearman's rank correlation coefficients for clicks and likes between the attributes.
As we confirmed the Pearson's correlation coefficients of the normalized values between attributes, we checked the correlation degree of the magnitude relation between articles using the rank correlation coefficient.
The like number's rank coefficient was weaker than that of the click number.
Especially for politics, the correlation was very weak in terms of age.
The reason that the rank correlation coefficient was smaller than the correlation coefficient of the normalized value for likes was that the rank of the articles changed significantly even if the number of likes changed only slightly.
For gender, the whole number of actions was divided into two attributes, whereas for age, the whole value was divided into three attributes; thus, the number of likes per age category was smaller than that per gender, and the rank coefficient for age was weaker.
As for the rank correlation coefficient, it is difficult to deal with small values, such as those for likes, so in the after section, Pearson's correlation coefficient is used for comparison.

\section{Comparison by Keyword}

In the previous section, differences in user behavior by attribute were compared using correlation coefficients and rank correlation coefficients.
In this section, we consider how the difference depends on the topic of the news article.
There are various definitions of news topics, but this study simply compares the articles based on the keywords included in the title.

\begin{figure}[tp]
  \centering
  \includegraphics[width=0.99\linewidth]{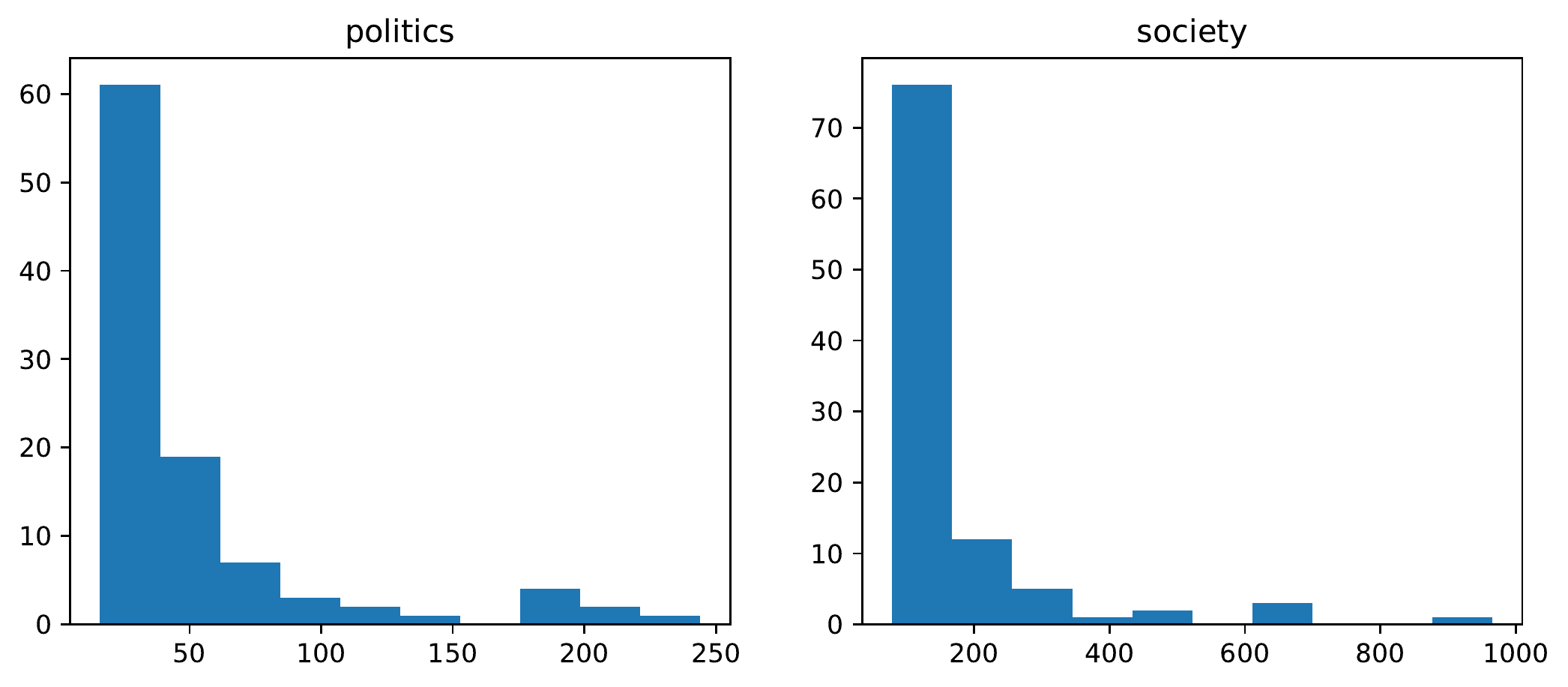}
    \caption{Distribution of the number of articles with each keyword present in the article's title.}
    \label{fig:word_distribution}
\end{figure}

First, we extract keywords from news articles. 
To confirm the correlation, the keyword needs to appear a certain number of times in datasets.
We divide the title of the news article into morphemes using morphological analysis, and these morphemes are taken as keyword candidates.
Then, the news articles including the keyword candidates are counted.
Following this, meaningless words are excluded manually, and we adopt the top 100 words in this count as keywords.
Fig.~\ref{fig:word_distribution} shows the distribution of frequency of the appearance of the selected keywords.

\begin{figure}[tp]
  \centering
  \includegraphics[width=0.99\linewidth]{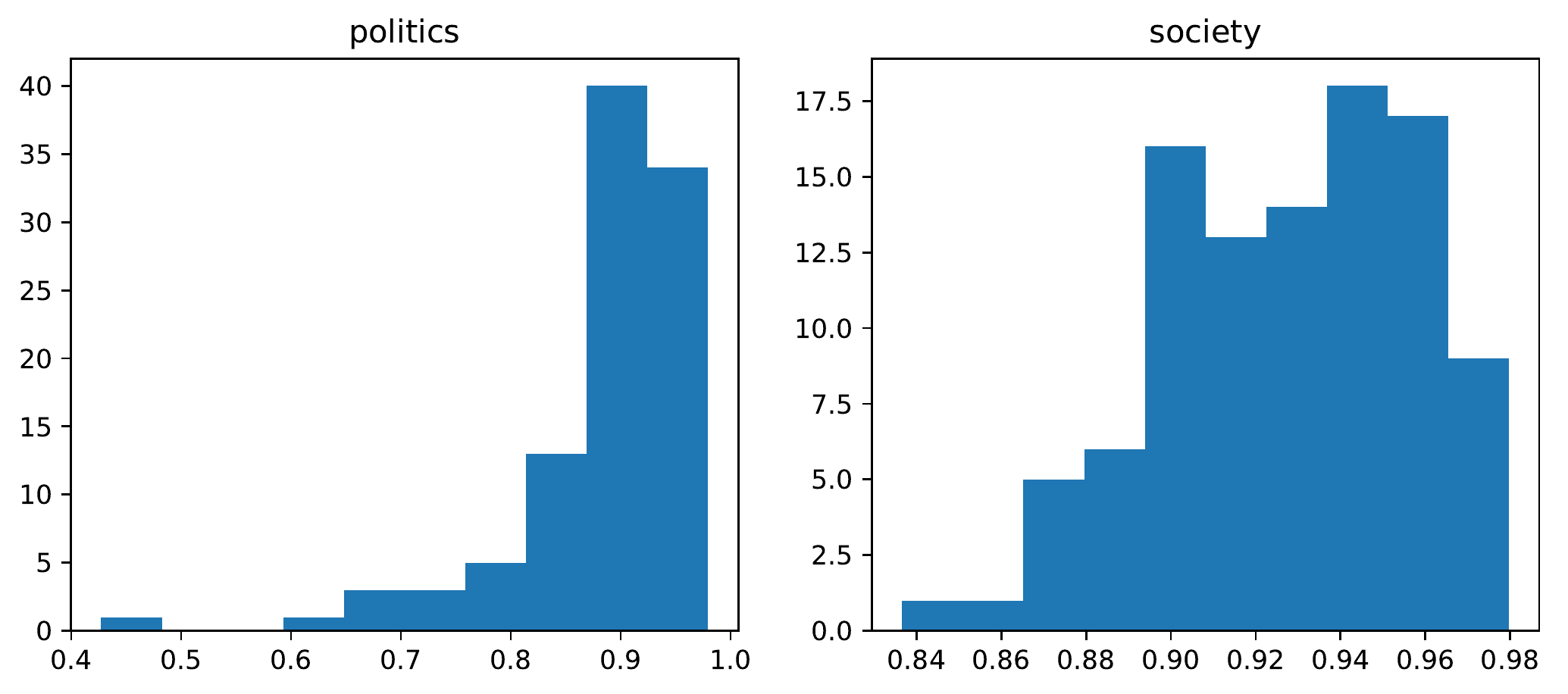}
    \caption{Distribution of keyword clicks' coefficient for gender.}
    \label{fig:keyword_clicks_gender_corr}
\end{figure}

\begin{figure}[tp]
  \centering
  \includegraphics[width=0.99\linewidth]{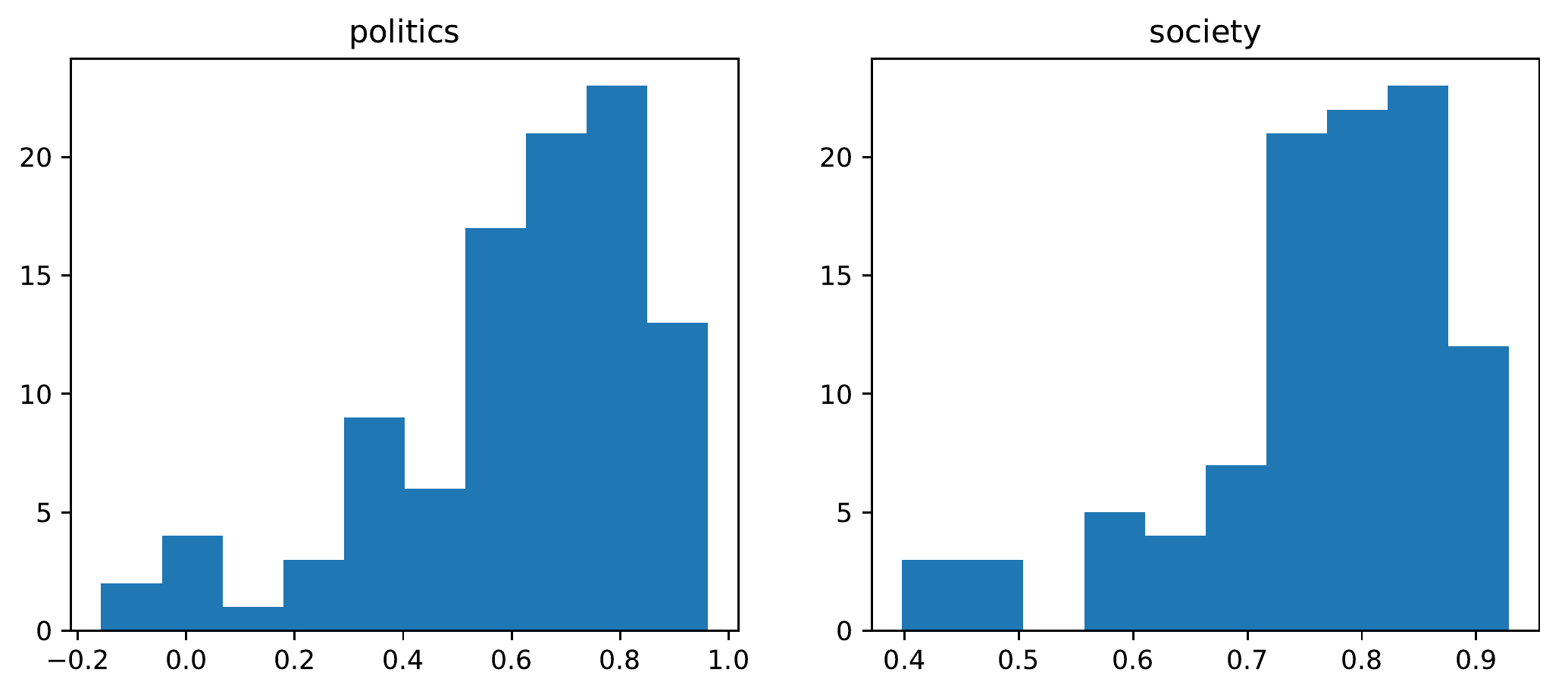}
    \caption{Distribution of keyword likes' coefficient for gender.}
    \label{fig:keyword_likes_gender_corr}
\end{figure}

\begin{figure}[tp]
  \centering
  \includegraphics[width=0.99\linewidth]{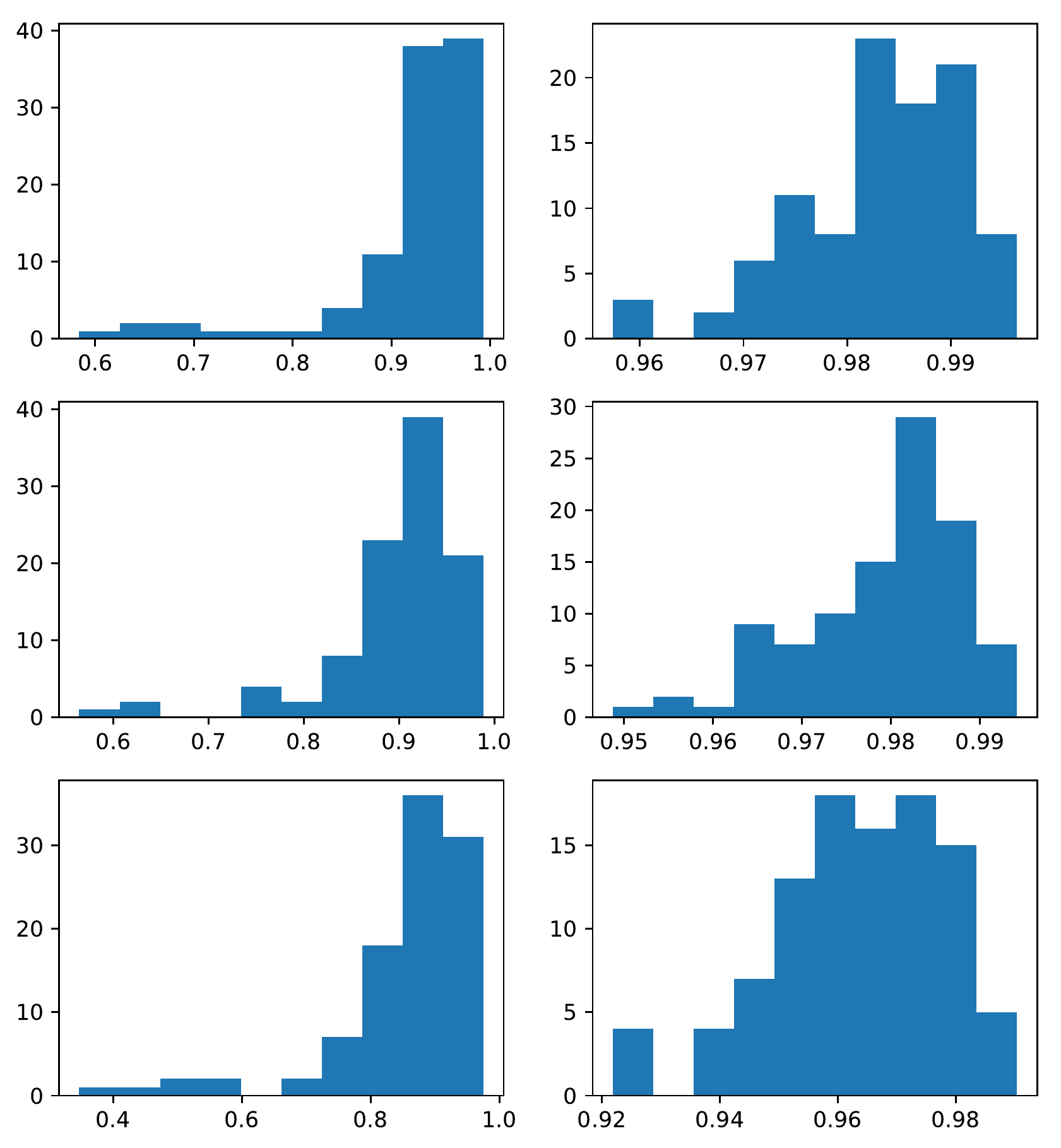}
    \caption{Distribution of keyword clicks' coefficients for age.}
    \label{fig:keyword_clicks_age_corr}
\end{figure}

\begin{figure}[tp]
  \centering
  \includegraphics[width=0.99\linewidth]{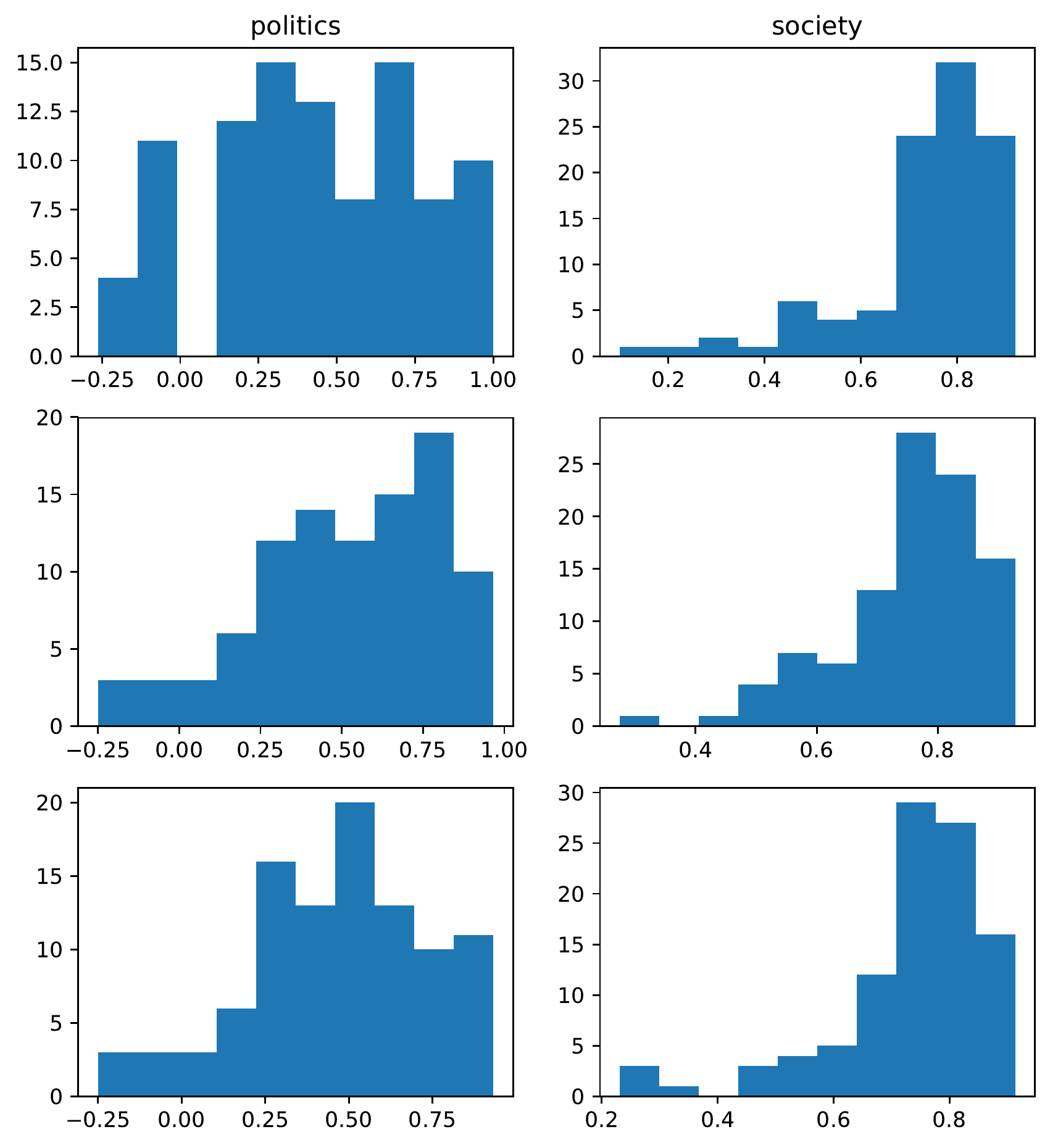}
    \caption{Distribution of keyword likes' coefficients for age.}
    \label{fig:keyword_likes_ager_corr}
\end{figure}

To confirm whether the comparison by keyword is valid, we confirm the correlation coefficient for each target keyword.
If the correlation of the click number or like number in the news articles between user attributes related to keywords is weak, there is no rule in the news articles related to keywords, so there is no point in comparing the keywords.
A news article containing the keyword in the title is taken as a news article on the keyword, and the correlation coefficient between the normalized click number and like number between the attributes is obtained, following which the distribution is confirmed.

The distribution of the correlation coefficients for each keyword is shown in Fig. \ref{fig:keyword_clicks_gender_corr} for the normalized number of clicks by gender, while Fig. \ref{fig:keyword_likes_gender_corr} shows the normalized number of likes by gender.
Furthermore, Fig. \ref{fig:keyword_clicks_age_corr} shows the normalized number of clicks by age, while Fig. \ref{fig:keyword_likes_ager_corr} shows the normalized number of likes by age.
Although there are some variations in the correlation coefficients between age in politics, the keywords have a high positive correlation between attributes.
In particular, there is a tendency for a higher correlation to be seen for society than for politics.
When the correlation coefficient is low, there is no characteristic tendency in the keyword.
Keywords with low correlation coefficients of likes in politics are those with articles with very few numbers of likes, so the correlation coefficient is considered to be low due to slight fluctuations.
The reason for this is that the correlation coefficient of age is lower than that of gender because of the different numbers of categories between the two attributes, as explained above.
From the above results, it was found that, if there were sufficient data, the approach seemed to explain the relationship between the attributes for each keyword.

Next, we run regression analysis for each keyword.
There was a strong correlation between attributes due to Pearson's correlation coefficient, but to identify biased keywords, we need to know how the behavior between attributes varies depending on the keywords.
In this study, our purpose is to uncover biased keywords by using regression analysis parameters.
In regression analysis of the keyword $k$ between attribute $A$ and attribute $B$, the normalized number of clicks of attribute $A$ of the article, including keyword $k$, is $y_{click, A}$, and the normalized number of clicks of attribute $B$ is $x_{click, B}$.
The regression curve is $y_{click, A}=a_{click, k, AB} x_{click, B} + b_{click, k, AB}$, and we can obtain the slope $a_{click, k, AB}$ and intercept $b_{click, k, AB}$.
For each keyword between gender and age, click and like regression analysis is performed, and each keyword is compared.
As already shown, some keywords have low correlation coefficients.
To properly compare them using regression analysis, keywords with a coefficient of determination of 0.5 or less in the regression analysis have been excluded from the comparison.

\begin{table}[tp]
\centering
\caption{Number of keywords in regression analysis. Keywords with coefficients of determination of 0.5 or less in the regression analysis were excluded.}
\scalebox{0.9}{
\begin{tabular}{|c|c|c|c|c|c|}
\hline
\multicolumn{2}{|c|}{\multirow{2}{*}{}} & \multicolumn{2}{c|}{Click} & \multicolumn{2}{c|}{Like} \\ \cline{3-6} 
\multicolumn{2}{|c|}{}                  & Politics     & Society     & Politics     & Society    \\ \hline
Gender                 & Male--Female    & 95           & 100         & 40           & 80         \\ \hline
\multirow{3}{*}{Age}   & Young--Middle   & 95           & 100         & 25           & 73         \\ \cline{2-6} 
                       & Middle--Older   & 97           & 100         & 33           & 77         \\ \cline{2-6} 
                       & Older--Young    & 93           & 100         & 22           & 74         \\ \hline
\end{tabular}
}
\label{tbl:n_keyword}
\end{table}

Table~\ref{tbl:n_keyword}. shows the number of target keywords.
The target keyword has fewer likes than clicks, and politics had fewer responses than society did.
This is the same trend as found for the correlation coefficient.

As a result of the regression analysis, we get intercept and slope values for each keyword.
Consider the meaning of this slice and slope.
\begin{figure}[tp]
  \centering
  \includegraphics[width=0.99\linewidth]{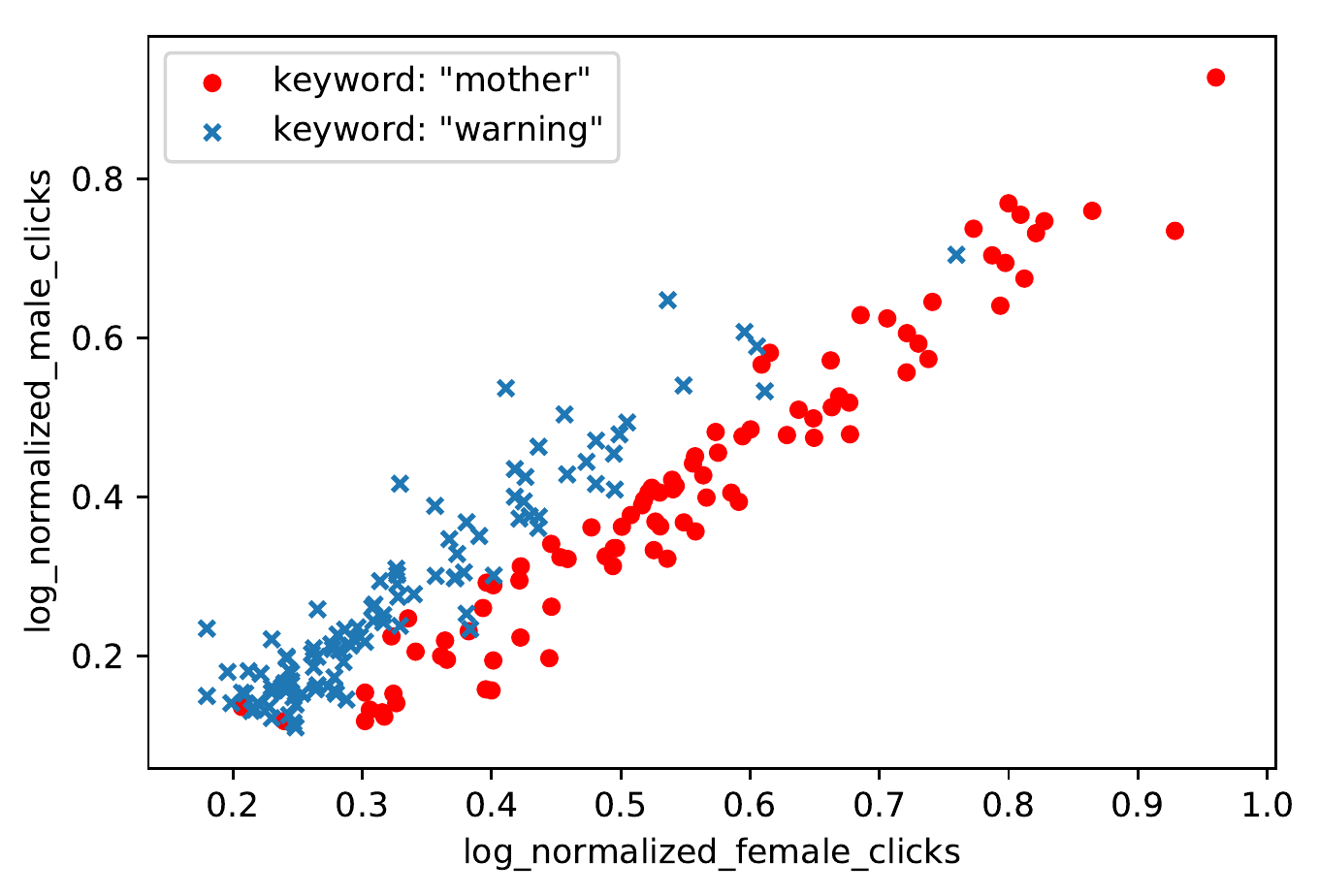}
    \caption{Scatter plots of two keywords of the normalized click number by gender for society. For {\it mother}, the intercept is very low and the slope is average, while for {\it warning}, the slope is the same and the intercept is not as low. }
    \label{fig:compare_intercept}
\end{figure}
First, we discuss the meaning of the intercept when the slope is about the same.
As an example, in Fig. \ref{fig:compare_intercept}, we show scatter plots of two keywords of normalized click numbers between genders for society.
For the first keyword, {\it mother}, the intercept is very low and the slope is average, while for the next keyword, {\it warning}, the slope is the same and the intercept is not as low.
Thus, the figure shows that the number of clicks of news articles including {\it mother}, with a small slice and comparable slope, is biased toward the female gender.

\begin{figure}[tp]
  \centering
  \includegraphics[width=0.99\linewidth]{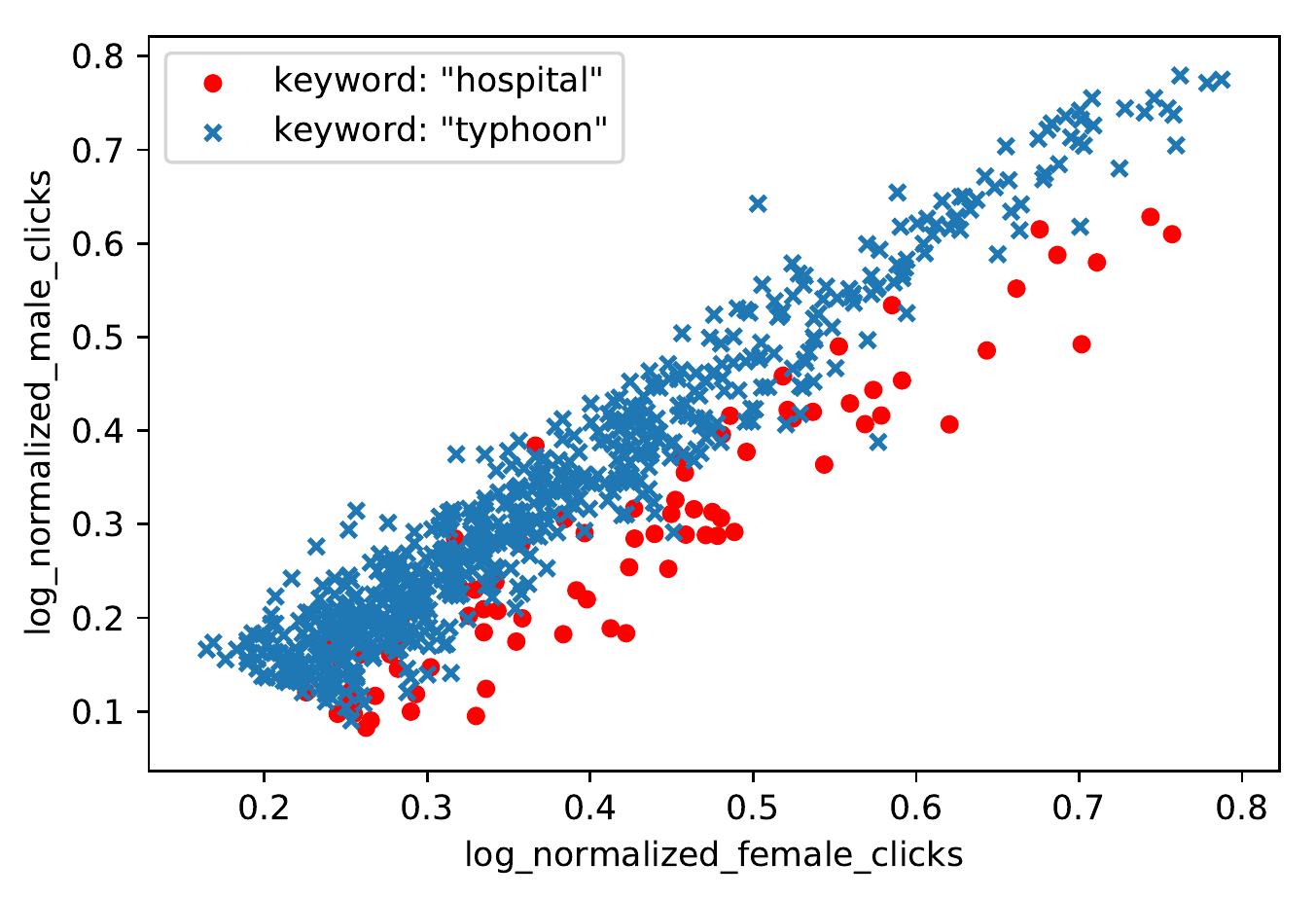}
    \caption{Scatter plots between males and females with the numbers of clicks of news articles including two keywords in society.
The first keyword is {\it typhoon}, for which the intercept is the average and the slope is large, and the second is {\it hospital}, which has a small intercept and gradient.}
    \label{fig:compare_coefient}
\end{figure}
Next, we discuss the meaning of the slope when the intercepts are of the same degree.
Similarly, as an example, we show scatter plots between males and females with the numbers of clicks of news articles including two keywords in society.
The first keyword is {\it typhoon}, for which the intercept is average and the slope is large; the second keyword is {\it hospital}, which has a small intercept and gradient.
As a result, news articles including {\it hospital} are often clicked on by females compared with news articles containing {\it typhoon}; thus, the keyword {\it hospital} is biased toward the female attribute.

If the intercept or slope of keywords is roughly the same, we can compare the behavior difference between the attributes using the other parameter.
Thus, we discuss cases where the intercept and slope are not the same by comparing {\it mother} and {\it hospital}, which are keywords that females are more likely to click on.
The slope of {\it mother} is 1.10, while the intercept is $-0.193$; in contrast, the slope of {\it hospital} is 0.983, and the intercept is $-0.125$. 
Thus, the values of parameters differ greatly.
\begin{figure}[tp]
  \centering
  \includegraphics[width=0.99\linewidth]{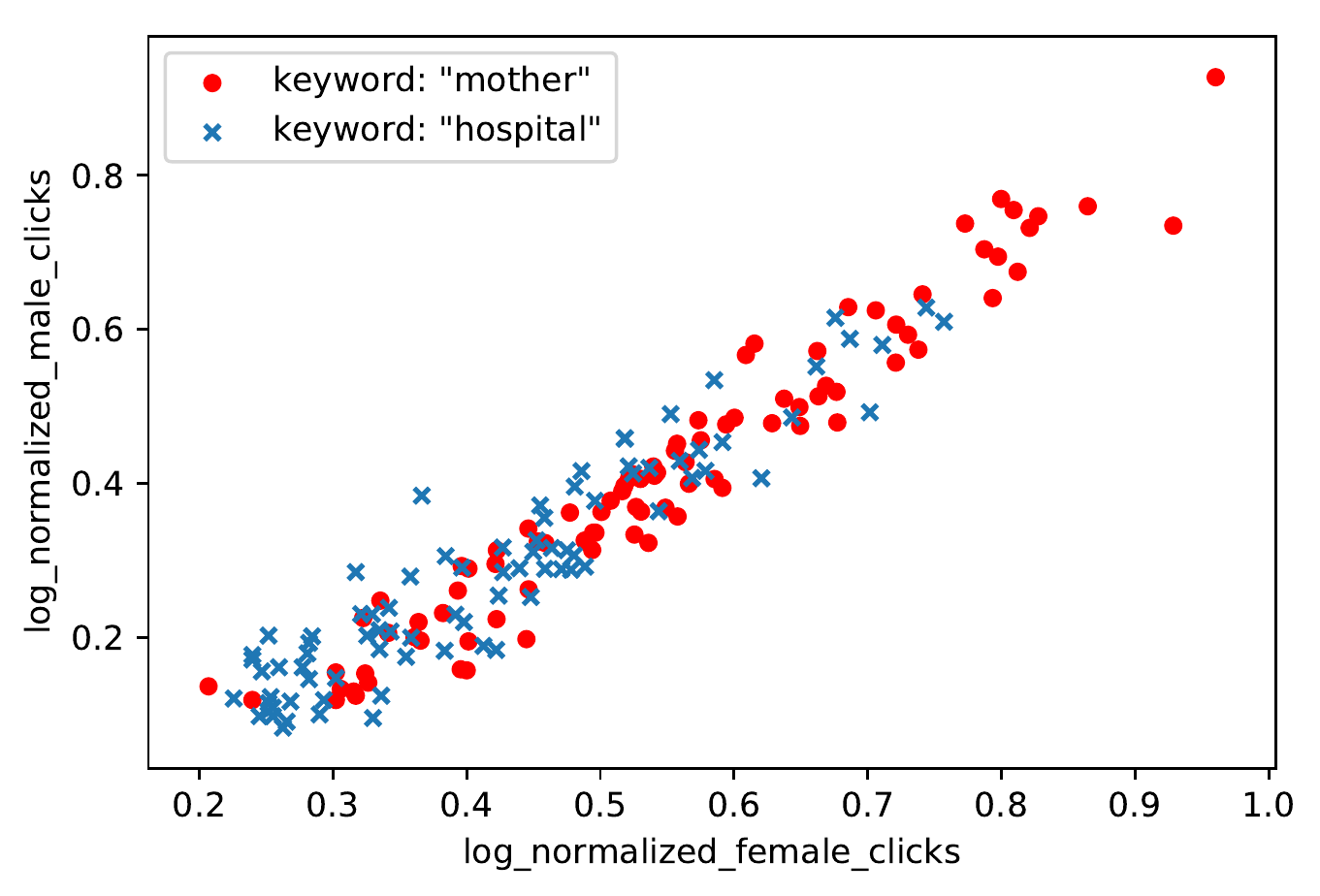}
    \caption{A two-keyword scatter plot distributed by gender in society: {\it mother} and {\it hospital} are shown as keywords in articles that females are likely to click on, as shown in Fig. \ref{fig:compare_intercept}, \ref{fig:compare_coefient}.}
    \label{fig:compare_female}
\end{figure}

Fig.~\ref{fig:compare_female} shows the gender distribution of clicks on news articles in the society category with the keywords of {\it mother} and {\it hospital}.
As shown in this figure, the click distribution between genders between the two keywords looks almost the same.
Despite this, there is a difference in parameters by the low number of clicks for {\it hospital} and high number for {\it mother}.
From the above, we are able to compare biases if the slope or intercept is the same, but it is difficult to compare them when both the slope and intercept are different.

\section{Discovering Bias Keywords}

In this study, the intercepts are divided into three classes, and in each class, the behavioral differences for the keywords are compared between attributes.
Based on the $ mean \pm \sigma \times 2$, the categories are classified as being within the section, larger than the upper end, and smaller than the lower end.\footnote{$\sigma$ is standard deviation}
The means of intercept classes are classified as being close to average, significantly larger, and significantly smaller.
Assuming that the distribution of intercepts of each keyword is a normal distribution, intercepts of keywords are classified as belonging to 95\% of the whole or not.
If the intercept is divergent from the mean, it can be seen that the behavior tendency is greatly biased to either attribute.
When the intercept is close to the mean, similarly to the intercept, keywords are extracted by the slope, which is deviated from the mean using standard deviation.

As shown in Fig. \ref{fig:compare_female}, it is difficult to determine which keyword is biased when the class is different, so this study does not compare keywords between classes, and this is left for future work.

\begin{table*}[tp]
\centering
\caption{List of biased keyword in class of upper or lower Intercept: When there is no keyword, a cell is blank. }
\scalebox{0.9}{
\begin{tabular}{|c|c|c|c|c|c|}
\hline
\multirow{2}{*}{}                                   & \multirow{2}{*}{Class}                                              & \multicolumn{2}{c|}{Politics}                                                                                                & \multicolumn{2}{c|}{Society}                                                                                                                                     \\ \cline{3-6} 
                                                    &                                                                     & Click                                                                                                         & Like         & Click                                                                                  & Like                                                                    \\ \hline
\multirow{2}{*}{Male--Female}                        & \begin{tabular}[c]{@{}l@{}}Upper \\ (biased to male)\end{tabular}   & House of Representatives, China                                                                               &              & Police                                                                                 & Obscenity                                                               \\ \cline{2-6} 
                                                    & \begin{tabular}[c]{@{}l@{}}Lower\\  (biased to female)\end{tabular} & \begin{tabular}[c]{@{}l@{}}Sugita, Mio Sugita, \\ Summer Time, Cabinet,\\ Olympics\end{tabular}               &              & Child, Mother                                                                          & \begin{tabular}[c]{@{}l@{}}Boy, Crash, Mother, \\ Children\end{tabular} \\ \hline
\multicolumn{1}{|c|}{\multirow{2}{*}{Young--Middle}} & \begin{tabular}[c]{@{}l@{}}Upper\\  (biased to young)\end{tabular}  & \begin{tabular}[c]{@{}l@{}}Constitutional change,\\ Olympics\end{tabular}                                     &              & \begin{tabular}[c]{@{}l@{}}Girl, University, \\ High school, Crash\end{tabular}        & Boy, Junior high school                                               \\ \cline{2-6} 
\multicolumn{1}{|c|}{}                              & \begin{tabular}[c]{@{}l@{}}Lower\\  (biased to middle)\end{tabular} & \begin{tabular}[c]{@{}l@{}}Candidacy announced, \\ Prefectoral assembly\end{tabular}                          & The Ministry & Boy                                                                                   & Incident, Murder, Search                                                \\ \hline
\multirow{2}{*}{Middle--Older}                       & \begin{tabular}[c]{@{}l@{}}Upper\\  (biased to middle)\end{tabular} & House of Representatives                                                                                      &              & Boy, Child, Tondabayashi                                                               & Bike                                                                    \\ \cline{2-6} 
                                                    & \begin{tabular}[c]{@{}l@{}}Lower\\  (biased to older)\end{tabular}  & \begin{tabular}[c]{@{}l@{}}Shigeru Ishiba, Takeshita Faction, \\ House of Representatives, China\end{tabular} & Mio Sugita,  & China                                                                                  & Mother, Boy                                                             \\ \hline
\multirow{2}{*}{Older--Young}                        & \begin{tabular}[c]{@{}l@{}}Upper \\ (biased to older)\end{tabular}  &                                                                                                               &              & China                                                                                  & Discover, Boy, Mother                                                   \\ \cline{2-6} 
                                                    & \begin{tabular}[c]{@{}l@{}}Lower\\  (biased to young)\end{tabular}  & Summer Time, Shinzo Abe                                                                                       &              & \begin{tabular}[c]{@{}l@{}}Girl,Boy, Tondabayashi,\\  High school, Child\end{tabular} &                                                                         \\ \hline
\end{tabular}
}
\label{tbl:biased_keyword_intercept}
\end{table*}

Table \ref{tbl:biased_keyword_intercept}. shows keywords with intercepts deviating from the mean.
A blank cell indicates that there was no keyword with a deviated intercept.
Since the purpose of this research is to extract these keywords, detailed discussion of meaning about each keyword is not done.
Several keywords indicate events in Japan in August 2018, which is the dataset period, so we describe its background.
The events related to the keywords that appeared in politics are shown below.
\begin{itemize}
    \item {\it Mio Sugita} is a Japanese politician who presented papers on LGBT in magazines. The claims in these papers is caused controversy.
    \item There is news about the possible introduction of Daylight Savings Time before the 2020 Summer Olympic Games in Tokyo.
    \item The election campaign for the leader of the LDP was held, matching Prime Minister {\it Shinzo Abe} against {\it Shigeru ishiba}.
\end{itemize}
Similarly, the events related to the keywords that appeared in politics are shown below.
\begin{itemize}
    \item A 2-year-old boy was missing in the forest and was rescued by a volunteer.
    \item A suspect escaped from a police station in {\it Toyodabayashi}.
\end{itemize}
Other keywords, such as mothers, girls, junior high school, and high school, were not connected to special big events.
There are many news articles including these keywords, not limited to that period, and it seems that there is a common bias in such news articles.

\begin{table*}[tp]
\caption{List of biased keyword in class of upper or lower slope: When there is no keyword, a cell is blank.}
\begin{center}
    \begin{tabular}{|c|c|c|c|c|c|}
\hline
\multirow{2}{*}{}             & \multirow{2}{*}{}                                                   & \multicolumn{2}{c|}{Politics}                                                                                             & \multicolumn{2}{c|}{Society}                                                                                                           \\ \cline{3-6} 
                              &                                                                     & Click                                                                                                   & Like            & Click                                                                 & Like                                                           \\ \hline
\multirow{2}{*}{Male--Female}  & \begin{tabular}[c]{@{}c@{}}Upper\\ (biased to male)\end{tabular}    & Provincial                                                                                              &                 &                                                                       &                                                                \\ \cline{2-6} 
                              & \begin{tabular}[c]{@{}c@{}}Lower\\ (biased to female)\end{tabular}  & Shinzo Abe                                                                                              & Shinzo Abe      & Search                                                                & \begin{tabular}[c]{@{}c@{}}Heatstroke,\\ Hospital\end{tabular} \\ \hline
\multirow{2}{*}{Young--Middle} & \begin{tabular}[c]{@{}c@{}}Upper\\ (biased to young)\end{tabular}   & Shinzo Abe                                                                                              & Yuichiro Tamaki &                                                                       & Gunma                                                          \\ \cline{2-6} 
                              & \begin{tabular}[c]{@{}c@{}}Lower\\ (biased to middle)\end{tabular}  & \begin{tabular}[c]{@{}c@{}}Denny Tamaki,\\ Takeshi Onaga\end{tabular}                                  &                 & \begin{tabular}[c]{@{}c@{}}Heavy rain,\\ Record,\\ House\end{tabular} &                                                                \\ \hline
\multirow{2}{*}{Middle--Older} & \begin{tabular}[c]{@{}c@{}}Upper\\ (biased to middle)\end{tabular}  & \begin{tabular}[c]{@{}c@{}}Mayor's election,\\ Candidacy\end{tabular}                                   &                 & \begin{tabular}[c]{@{}c@{}}Weather,\\ Obon,\\ Search\end{tabular}     & Girl                                                           \\ \cline{2-6} 
                              & \begin{tabular}[c]{@{}c@{}}lower\\ (biased to older)\end{tabular}   & \begin{tabular}[c]{@{}c@{}}Shinzo Abe\\ Minister of Internal Affairs \\ and Communications\end{tabular} &                 & Fukuoka                                                               & Typhoon No.19                                                  \\ \hline
\multirow{2}{*}{Older--Young}  & \begin{tabular}[c]{@{}c@{}}Upper\\  (biased to older)\end{tabular}  & \begin{tabular}[c]{@{}c@{}}Minister of Internal Affairs \\ and Communications\end{tabular}             &                 &                                                                       & \begin{tabular}[c]{@{}c@{}}China,\\ Search\end{tabular}        \\ \cline{2-6} 
                              & \begin{tabular}[c]{@{}c@{}}Lower\\ (biased to young)\}\end{tabular} & \begin{tabular}[c]{@{}c@{}}Shinzo Abe,\\ Yuichiro Tamaki\end{tabular}                                   &                 & \begin{tabular}[c]{@{}c@{}}Kyoto,\\ Earthquake,\\ Obon\end{tabular}   & Inspection                                                     \\ \hline
\end{tabular}
\end{center}
\label{tbl:biased_keyword_slope}
\end{table*}

Table~\ref{tbl:biased_keyword_slope}. shows the keywords whose slope deviates from the average among the keywords whose intercepts are close to average.
The events related to keywords that appeared in the politics are as follows.
\begin{itemize}
    \item In the election race for the leader of the LDP mentioned above, it was reported that {\it Seiko Noda}, Minister of Internal Affairs and Communications, would also run as a candidate.
    \item The National Democratic Party's representative election was held; {\it Yuichiro Tamaki} have ran as a candidate.
    \item {\it Okinawa} Governor {\t Onaga} suddenly died, the {\it Okinawa} prefectural governor's election was held, and {\it Denny Tamaki} have ran as a candidate. And the {\it Ginowan} mayor's election was held on the same day.
\end{itemize}
The events related to keywords that appeared in society are as follows.
\begin{itemize}
    \item Many people died of heatstroke due to the hot weather.
    \item Record-setting heavy rain hit West Japan due to the influence of Typhoon 7.
    \item Serious typhoons 19 and 20 approached the Japanese archipelago.
\end{itemize}
Prefecture names such as Fukuoka, Gunma, and Kyoto appeared, but there was not any big event in particular.
{\it Obon} is a Japanese traditional festival like Halloween.
Some earthquakes have occurred in various places, but a big one occurred on Lombok Island in Indonesia.
Also, the news that a man who was working on the removal of the Osaka North Earthquake died has been widely reported.

As described above, we were able to extract keywords that are thought to be biased among attributes from parameters of regression analysis deviating from the mean.
Some keywords are difficult to link with clear topics, but many keywords are related to events that occurred during the period, and we discuss the reaction of each user attribute to various events from this list.

\section{Conclusion}

In this paper, we analyzed behavior differences between user attributes based on the user behavior log of news applications and extracted keywords with biased behavior.
The purpose of this research is to clarify behavior bias in users' news applications.
First, we show that behavior among users' attributes varies in categories, indicating that there is a difference in behavior between attributes.
Then, we show that the number of clicks of the news articles normalized in the attribute and the number of likes are strongly correlated among the attributes, and it is shown that the differences between the attributes can be compared.
Then, we found correlation between attributes for each keyword included in the title of the news article, and since there are many keywords with strong correlation, we confirmed that we can compare the differences between attributes by keywords.
Regression analysis was performed on each behavior between attributes for each keyword, and it was possible to obtain a biased keyword from the degree of departure from the average value of slope and intercept.

Finally, we discuss future works.
In this paper, we do not discuss the validity of biased keywords.
We would like to aim to verify whether this result is valid according to social science knowledge.
Also, we plan to discover a strong bias topic due to clusters of user's interests, etc., rather than user attributes.
Regarding extraction of keywords by regression analysis, processing of parameters is complicated.
We will create a measure that can extract keywords more simply.

\bibliographystyle{IEEEtran}
\bibliography{IEEEabrv,mybib}

\end{document}